\documentclass[sigconf,table]{acmart}

\renewcommand\footnotetextcopyrightpermission[1]{} %
\usepackage{enumitem}
\usepackage{graphicx}
\usepackage{hyperref}
\usepackage{url}
\usepackage{multirow}

\usepackage{color, colortbl}
\usepackage[pscoord]{eso-pic}%

\newcommand{\preprintBanner}{
\AddToShipoutPictureFG*{\put(\LenToUnit{0.5\paperwidth},\LenToUnit{0.95\paperheight}){\makebox[0pt][c]{
\renewcommand{\arraystretch}{1.5}
\setlength{\tabcolsep}{18pt}
\rowcolors{1}{Preprint}{Gray}
\begin{tabular}{|p{0.55\paperwidth}|} \hline
\textbf{Preprint from \texttt{\href{https://ostendorff.org/pub/}{https://ostendorff.org/pub/}}} \\ \hline
\footnotesize
M. Ostendorff ``Contextual Document Similarity for Content-based Literature Recommender Systems'' in \textit{Proceedings of Doctoral Consortium at the ACM/IEEE Joint Conference on Digital Libraries (JCDL)}, 2020. \\ \hline
\end{tabular}}}}
}

\definecolor{Gray}{gray}{0.925} %
\definecolor{Preprint}{rgb}{.63,.79,.95}

\copyrightyear{2020}

\acmConference[]{}
\acmBooktitle{}

\begin{document}

\fancyhead{}

\title{Contextual Document Similarity for Content-based Literature Recommender Systems}

\author{Malte Ostendorff}
\affiliation{
  \institution{University of Konstanz}
  \city{Konstanz}
  \country{Germany}
}
\email{malte.ostendorff@uni-konstanz.de}

\affiliation{~}
\affiliation{Doctoral Advisor: Prof. Dr. Bela Gipp}

\begin{abstract}

To cope with the ever-growing information overload, an increasing number of digital libraries employ content-based recommender systems.
These systems traditionally recommend related documents with the help of similarity measures.
However, current document similarity measures simply distinguish between similar and dissimilar documents. 
This simplification is especially crucial for extensive documents, which cover various facets of a topic and are often found in digital libraries.
Still, these similarity measures neglect to what facet the similarity relates.
Therefore, the context of the similarity remains ill-defined.
In this doctoral thesis, we explore contextual document similarity measures, i.e., methods that determine document similarity as a triple of two documents and the context of their similarity.
The context is here a further specification of the similarity. 
For example, in the scientific domain, research papers can be similar with respect to their background, methodology, or findings.
The measurement of similarity in regards to one or more given contexts will enhance recommender systems. 
Namely, users will be able to explore document collections by formulating queries in terms of documents and their contextual similarities. 
Thus, our research objective is the development and evaluation of a recommender system based on contextual similarity.
The underlying techniques will apply established similarity measures and as well as neural approaches, while utilizing semantic features obtained from links between documents and their text.

\end{abstract}



\maketitle

\preprintBanner

\section{Introduction}
\label{sec:intro}

\begin{figure}[h]
\centering
\includegraphics[page=16,clip,width=0.4\textwidth,trim={1cm 1.5cm 1cm 1.5cm}]{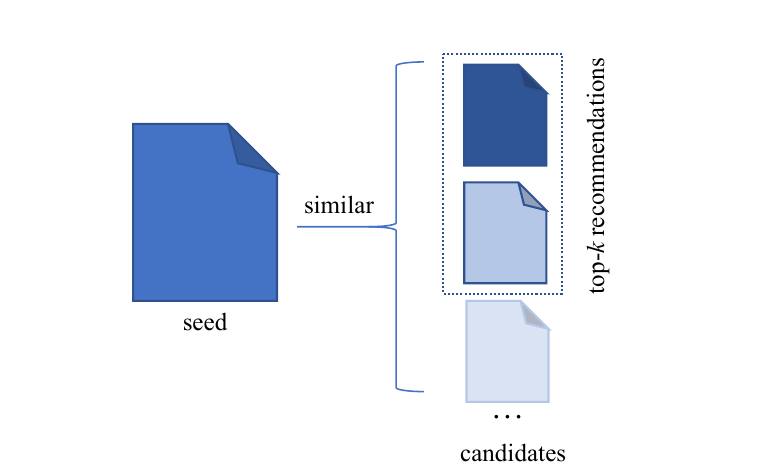}
\caption{\label{fig:1d-similarity}Traditional similarity: Document semantics are assumed to be unambiguous. Recommendations are chosen among candidate documents based on a single similarity measure.}
\end{figure}

Recommender systems are a popular filtering and discovery tool for managing the vast and continuously increasing amount of digitally available content. 
Many systems, like collaborative filtering~\cite{Schafer2007}, gather information about their users and provide individual recommendations based on the collected data.
However, in numerous scenarios, user-based recommender systems are not applicable, e.g., due to privacy constraints or too frequent changes in the user's information need to provide meaningful recommendations.
Digital libraries often deal with such a scenario.
Instead of user-based features, most of the literature recommender systems (approximately 55\%) employ content-based document features and corresponding similarity measures~\cite{Beel2016}.
The task of recommending documents is often divided into two major phases, feature representation, and retrieval. 
First, features of documents are represented as numerical vectors, both the query document (seed) and document collection. 
Translating words and documents into n-dimensional vectors is a common task in information retrieval (IR) and natural language processing (NLP).
The vector space model~\cite{Salton1975}, TF-IDF~\cite{Jones1972}, Paragraph Vectors~\cite{Le2014}, and GloVe~\cite{Pennington2014} among others have been proven to be effective to capture semantic text features. %
Similarly, non-textual document elements, like citations used in scientific literature or hyperlinks in web pages, are an essential source of semantic information~\cite{Kessler1963,Small1973,Garfield2001,Gipp2009}. %
Second, a retrieval method selects documents from the collection that are most similar to the seed document.
The cosine similarity is one common measure that computes the similarity score between document vectors.
As illustrated in Figure~\ref{fig:1d-similarity}, a (one dimensional) similarity score is assigned to each document pair of seed and recommendation candidate. 
Then, the top-$k$ recommendations are chosen from the candidate documents with the highest similarity to the seed document.

However, this approach does not account for the unambiguous semantics of long-form documents.
For instance, research papers tend to cover multiple facets of a topic, e.g., method, background, or results.
Still, today's similarity measures treat documents as singular entities even tough the document semantics are rather heterogeneous.
As a result, it remains unclear to what facets the similarity rates.
Philosopher Nelson Goodman~\cite{Goodman1972} already argued that the similarity of A to B is a meaningless notion unless one can say ``in what respect'' A is similar to B.
In the context of NLP, B\"{a}r et al.~\cite{Bar2011} found that \textit{text similarity} is used indiscriminately as a general term without any concern of the many perspectives, two different item can be co-related. 
Consequently, the similarity of A and B should only make sense if we know what context is taken into account. 

\begin{figure}[h]
\centering
\includegraphics[page=15,clip,width=0.4\textwidth,trim={1cm 1.5cm 1cm 1cm}]{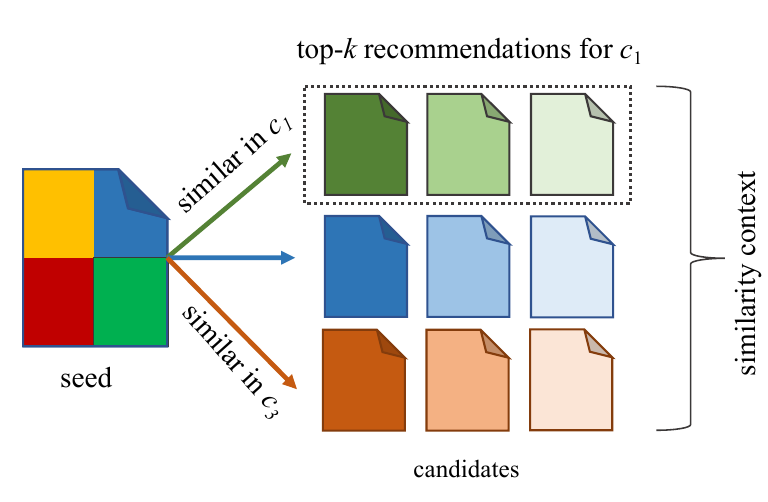}
\caption{\label{fig:contextual-similarity}Contextual document similarity: Document semantics are ambiguous. Multiple similarity dimensions reflect the different aspects of a document (illustrated by different colors).}
\end{figure}

Figure~\ref{fig:contextual-similarity} depicts how the similarity of documents can change depending on a given context.
For context $c_1$ (green), the most similar document is different from the one for context $c_3$ (orange).
In practise, the different similarity contexts could be, for example, the background, methodology, or findings of research papers when working with scientific literature.
Determining the contextual document similarity would be, in particular, beneficial for recommender systems with an expert audience.
Experts have complex information needs and often search for relations between literature that may not be evident at first sight.
For the domain of scientific literature, Chan et al.~\cite{Chan2018} emphasize how important the discovery of analogies between research papers is for scientific progress. 
Current document similarity measures are not designed for analogical queries~\cite{Gick1983}.
An example of an analogical query would be the retrieval of other papers with similar methodology but different outcome.
Solving such queries is crucial for finding distantly related, yet, highly relevant documents.
One relevant paper that shares a specific aspect might remain undiscovered because it is from a different research field and, therefore, does not share the vocabulary nor citations with the seed document. 

In this thesis, we will investigate what we define as contextual document similarity. 
Instead of determining similarity as a tuple of a seed document $d_s$ and a target document $d_t$, the contextual document similarity is a triple of $d_s$, $d_t$ and a context $c_i$.
As the source of information for the context, we focus on textual content and links (or citations) between documents. 
We hypothesize that similarity-based recommender systems can be significantly improved by considering the context of the similarity.
Moreover, we envision improvements for systems with expert users, who benefit from recommendations without any evident connection to the seed document.

\section{Research Approach}
\label{sec:approach}

In contrast to established document similarity measures, the contextual document similarity will account for many-faceted document semantics, and will clarify in what respects two documents are similar or dissimilar.
Downstream information retrieval applications will benefit from the context information.
To be precise, we focus on literature recommender systems as application.

\subsection{Research Objective} \label{ssec:objectives}

The shortcomings of document similarity measures and content-based recommender systems motivate us to define the following research objective:
\\[12pt]
\textit{Design, implement, and evaluate a contextual document similarity measure that utilizes link- and text-features to enable recommendations of relevant literature, which would remain undiscovered with existing methods.}
\\[12pt]

The goal of the planned research project is twofold.
First, we will conceive a contextual document similarity measure that utilizes hybrid semantic features from text and links, whereby the measure provides the context in which two documents are similar.
Here, the text is a natural language text from the document body and links are structural elements, like citations, that connect two documents.
Two documents $d_s$ and $d_t$, and a third element, the context $c_i$, define the contextual document similarity.
The further specification of the context $c_i$ is subject to this research.
Second, we will develop a recommender system that allows the exploration of the document space in terms of documents and their relations. 
The system will enable analogical queries, similar to SPARQL queries, and will be empirically tested in a user-centric evaluation.
Based on a given seed document, the system finds other documents that share one or more similarity contexts with the seed document or miss one or more other similarity contexts.

\subsection{Research Tasks and Questions} \label{ssec:research-tasks}

To achieve this objective, we divide the thesis into the following research tasks:

\begin{itemize}

\item[\textbf{T1}] Review the strengths and weakness of state-of-the-art document representations and the corresponding similarity measures with the focus on answering:

\begin{itemize}

\item[a)]{How do today's representation techniques account for the many facets of extensive documents? }

\item[b)]{What methods exist to jointly utilize text and link-based features for a single similarity measure?}

\item[c)]{What are the requirements for a contextual similarity measure that can express to what the similarity of two documents relates?}

\item[d)]{What should be defined as the similarity contexts to improve literature recommendations?}

\end{itemize}

\item[\textbf{T2}] Design and implement a novel method that addresses weakness and combines the strengths of existing methods, whereby semantic features for text and links should be combined in a hybrid manner.

\item[\textbf{T3}] Develop a prototypical system that leverages the contextual document similarity measure for enhanced literature recommendations.

\item[\textbf{T4}] Conduct a user-centric evaluation comparing the prototype with existing systems in regards to it performance (recommendation quality, satisfying information need), whereby the discovery of distantly related, yet, relevant documents is the primary objective.

\end{itemize}

This final step of the recommender system evaluation reflects the fact that contextual document similarity does not serve a purpose on its own.
Instead it is only the underlying technology that ultimately aims to enable users to make sense of a large document collection.
Whether we achieve this goal can consequently only be proven by exposing the prototypical system to users in a real-world application, and examining the users' behavior.

\section{Related Work} \label{sec:sota}

\subsection{Text-based Similarity}

A classical method for representing natural language text as a numerical vector is bag-of-words~\cite{Harris1954} or the vector space model~\cite{Salton1975}. 
Both methods produce sparse vector representations of documents in which the dimensions correspond to the terms used in the document corpus, and the values indicate how many times a term occurred in a document.
With Term Frequency - Inverse Document Frequency~\cite{Jones1972} (TF-IDF), the values depend on the term specificity, which improve the semantic document representation.
In combination with cosine similarity, sparse vector representations allow the efficient computation of document similarities, are integrated into popular information retrieval frameworks, e.g., Apache Lucene\footnote{\url{https://lucene.apache.org}}, and have been successfully tested in recommender system research~\cite{Collins2019,Kanakia2019,Schwarzer2016}.

With word2vec, Mikolov et al.~\cite{Mikolov2013} introduced an algorithm to learn dense vector representations of words such that semantically similar words end up close to each other in the embedding space. 
Word2vec has two different training algorithms, named continuous bag-of-words (cbow), and skip-gram. The former predicts a word based on its context, and the latter the converse. Both models, cbow and skip-gram, are widely applied in NLP tasks~\cite{Iacobacci2016,Ruas2019}, but they do not represent entire documents. 
The concept of Paragraph Vectors~\cite{Le2014} extends word2vec to learn embeddings for word sequences of arbitrary length.
Paragraph Vectors also has two training algorithms, named distributed memory (dm) and distributed bag-of-words (dbow), which are analogous to cbow and skip-gram, respectively. 
In both algorithms, Paragraph Vectors uses an extra vector responsible for capturing the semantic representation of the entire text document. 
The dbow training model has been shown to outperform dm in semantic similarity tasks~\cite{Lau2016}.
Aside from word2vec, other prominent techniques for word embeddings are GloVe~\cite{Pennington2014} and fastText~\cite{Bojanowski2017}.
Arora et al.~\cite{Arora2017} showed that taking the weighted average of word vectors is  a simple, yet effective method to derive document representations, and can outperform more sophisticated methods.

More recently, Transformer-based~\cite{Vaswani2017} language models introduced a shift from context-free word embeddings, like word2vec~\cite{Mikolov2013} or GloVe~\cite{Pennington2014}, to contextual embeddings as used in BERT~\cite{Devlin2019}, RoBERTa~\cite{Liu2019}, Transformer-XL~\cite{Dai2019}, XLNet~\cite{Yang2019} and others.
The Transformer architecture allowed the large-scale unsupervised pretraining of language models and led to significant improvements for many NLP tasks.
Reimers et al.~\cite{Reimers2019} show that BERT can be successfully applied to determine sentence similarity.
In prior work, we also applied BERT successfully for document classification~\cite{Ostendorff2019}.
In contrast to the rather classical methods of sparse document vector or context-free word embeddings, Transformers are computationally more expensive, especially on longer text sequences.  
Until now, Transformers gained little attention in the recommender system community. 
This is presumably due to the computation resources that Transformers require. 
A recent recommender system survey by Bai et al.~\cite{Bai2019} does not report by any use of Transformers. 
To our knowledge, Hassan et al.~\cite{Hassan2019} are one of the first to use BERT to recommend research papers, whereby they only use BERT to encode only the paper titles and not the document text as vectors .

\subsection{Link-based Similarity}

Besides the document text, links between documents supply essential semantic information. 
We refer to links for any kind of structural elements that connect documents, that can be citations in academic papers, and also hyperlinks in web pages.
Already in ``pre-Web'' times, links in the form of citations played a crucial role in library science.
In 1963, Kessler~\cite{Kessler1963} introduced Bibliographic Coupling to determine a similarity relationship between two documents depending on the overlap of their cited literature. 
Small~\cite{Small1973} presented Co-citation as the frequency with which two documents are cited together by other documents.
The more often two documents are co-cited together, the degree of similarity increases.
In contrast to Bibliographic Coupling, Co-citation relies on external factors and changes over time~\cite{Garfield2001}.
With the increasing availability of full-text publications, it became evident to also utilize the position citations within the text for similarity assessments.
Gipp and Beel~\cite{Gipp2009} incorporated the citation position as a semantic feature into their Co-citation proximity analysis (CPA).
The underlying idea of CPA is that, when citation markers of co-cited documents are in close proximity, the documents are more likely to be similar.
Empirical studies show the positive effect of co-citation proximity~\cite{Eto2013,Schwarzer2016,Knoth2017}.

Aside from the position, the textual context in which a link or citation occurs is another valuable semantic information.
In scientific literature, the text around citations tends to state known facts, similar to summaries, more concisely than in the original papers.
Nakov et al.~\cite{Nakov2004} demonstrate that citation context information aligns well with manually curated information from the biomedical domain.
At the same time, the citation context also depends on the citing paper. 
Elkiss et al.~\cite{Elkiss2008} found that different citations to the same paper often focus on different aspects of that paper.
As a result, we plan to exploit the information implied in the citation context to derive a context for the document similarity of the cited and the citing document.
Moreover, the availability of citation context datasets~\cite{Cohan2019,Jurgens2018,Athar2012} might also allow developing a learning approach for documents that do not share any citations. 

To easily integrate link-based features into a learning approach, one needs to derive vector representations.
Documents connected by links are essentially a graph and, therefore, methods for generated graph embeddings are also applicable.
Graph embedding techniques are closely related to the ones used for word embeddings.
Methods like DeepWalk~\cite{Perozzi2014} or Node2vec~\cite{Grover2016}, learn node embeddings by leveraging the local structures in the network similar to word2vec~\cite{Mikolov2013}. 
The local structure in a graph corresponds to the words occurring in the same sentence.
Various studies demonstrate the application of graph embedding for linked documents~\cite{Berger2017,Ganguly2017,Han2018}.
However, to our knowledge, none of these studies take the link proximity as in CPA~\cite{Gipp2009} into account.
Thus, we will investigate CPA's transferability to graph embeddings.

\subsection{Recommender Systems}

Recommender systems are often divided into four categories~\cite{Jannach2010}: collaborative filtering, knowledge-based, content-based and hybrid recommender system.
In this work, we primarily focus on content-based recommender systems, whereby content referrers to any features that originate from the recommended items.
As application domain, we focus on academic literature, which is also subject to related surveys~\cite{Beel2016,Bai2019}. 
Kanakia et al.~\cite{Kanakia2019} use the Microsoft Academic Graph~\cite{Farber2019} to build a recommender system for research papers. They construct two recommendation sets, one with Co-citation~\cite{Small1973} and another one with text-based features from TF-IDF~\cite{Jones1972} and word2vec~\cite{Mikolov2013}, and combine the two sets in a hybrid manner.
In comparison to systems only based on text or citation features,  Kanakia et al. find that their hybrid approach correlates stronger with the relevance judgments of 40 user study participants. %
This outcome motivates us to as well make use of hybrid features.
Collins and Beel~\cite{Collins2019} conduct an online evaluation of three text-based using TF-IDF~\cite{Jones1972}, Paragraph-Vectors~\cite{Le2014}, and Keyphrases~\cite{Ferrara2011}. 
They serve 33.5M recommendations to users of two digital libraries and measure the performance in terms of click-through rate (CTR). 
Collins and Beel observe a significant performance difference between the algorithm (up to ~400\%) but also between the two libraries.
This outcome highlights the importance of recommender system research and especially the challenges of evaluations.
Experiments might not be reproducible~\cite{Konstan2013,Beel2016b}, or metrics like CTR might not correspond to actual relevance~\cite{Zheng2010}.

Therefore, the user-centric evaluation will substantial to this research.
Betts et al.~\cite{Betts2019} demonstrate that, aside from citations and text, additional metadata (authors, affiliations, venues) allows them to construct a searchable graph database for scientific literature.
With contextual document similarity, we aim to achieve similar graph-like structures but based on the semantic similarity of the document content.

\section{Towards Contextual Document Similarity for Literature Recommendations}

In the first stage of the doctoral thesis, we will review, combine, and adapt existing techniques for document representations and measures to define their contextual similarity (T1).
On the one hand, we must consider the recent advances in neural methods. 
In particular, techniques that been shown to be useful in other fields, e.g., NLP tasks, but have not yet been applied for recommendations, are of interest.
One example of such a technology would be the Transformer~\cite{Vaswani2017}.
On the other hand, rather classical but well-established methods, such as CPA~\cite{Gipp2009}, which do not have their neural counterpart,  must not be neglected.
Moreover, hybrid techniques, as shown by Kanakia et al.~\cite{Kanakia2019}, can combine strengths from links and text-based approaches. 
Accordingly, text and link combination will be likewise in our focus.

\begin{figure}[h]
\centering
\includegraphics[page=17,clip,width=0.45\textwidth,trim={3.3cm 2cm 3.5cm 2cm}]{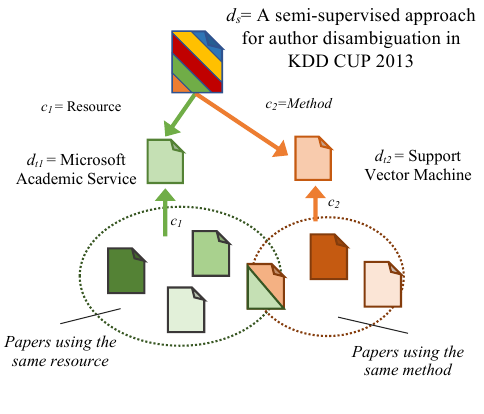}
\caption{\label{fig:paper-example}Contextual document similarity for research papers. The seed paper \cite{Zhao2013} is similar with respect to its resource (green) or method (orange) to other papers.}

\end{figure}

\begin{figure*}[t]
\centering
\includegraphics[clip,width=0.95\textwidth]{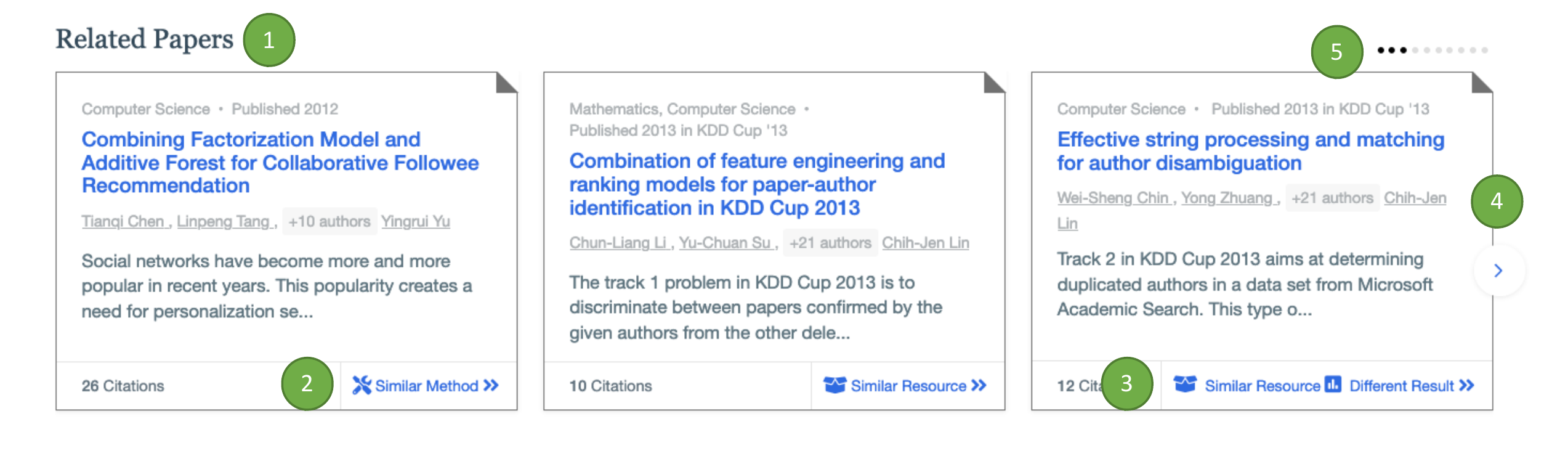}
\caption{\label{fig:recsys-example}Example of context information being integrated into the Semantic Scholar recommendation interface. 
1) Recommendations are listed in a ``Related Papers'' section. 
2) Each recommendation is accompanied by context information, e.g., similar method. Users can click on the icon to browse more papers focused on a particular context.
3) Combination of similar and different contexts.
4) Users can navigate to additional recommendation sets with other context combinations by clicking on the arrow.
5) The dots indicate the availability of other recommendations.  }
\end{figure*}

Based on the findings from the literature review, we will conceive a contextual document similarity measure to address the shortcoming of existing methods (T2).
Current document similarity measures usually define the similarity of a seed document $d_s$ and target document $d_t$ in terms of a single similarity score, e.g., cosine similarity of document vectors or co-citation proximity index for CPA~\cite{Gipp2009} (Figure~\ref{fig:1d-similarity}).
These similarity scores are scalar values and usually normalized to the range from 0 (dissimilar) to 1 (similar). 
The contextual similarity measure goes one step further and is illustrated in Figure~\ref{fig:contextual-similarity}.
To be precise, the contextual similarity will measure the similarity of $d_s$ and $d_t$ with respect to a given context $c_i$.
Therefore, the contextual similarity will be computed as a function $\text{sim}(d_s,d_t,c_i)=[0,1]$.
The contexts are denoted as the finite set $C=\{c_1,c_2,c_3,\ldots,c_n\}$ and depend on the document corpus and use case.
Even if the further specification of $C$ is part of this research, Figure~\ref{fig:paper-example} depicts one example for recommendations of research papers.
In the example, there are two similarity contexts in regards to resource $c_1$ and method $c_2$ used in the paper.
The paper by Zhao et al.~\cite{Zhao2013} on author disambiguation exemplifies the seed document.
Support Vector Machine~\cite{Cortes1995} is similar to the seed concerning method $c_2$, while the Microsoft Academic Service~\cite{Farber2019} relates to the resource context $c_1$.
The example's underlying information originates from \cite{Zhao2013}'s entry\footnote{\url{https://www.orkg.org/orkg/paper/R6119/R6120}} in the Open Research Knowledge Graph~\cite{Jaradeh2019}.
The contextual similarity measure aims for an outcome comparable to the one in Figure~\ref{fig:paper-example}.

Such an outcome will pave the way for the last stage of the research that is the development (T3) and evaluation (T4) of a recommender system based on contextual document similarity.
We envision mainly two opportunities to advance recommender systems, (i) diverse and (ii) focused recommendations. 
As the context describe different facets of the seed document, one could diversify the recommendations. 
Choosing the recommendations from documents that are similar in different contexts to the seed document would ensure diversity. 
In Figure~\ref{fig:paper-example}, \cite{Cortes1995} and \cite{Farber2019} can be considered as diverse recommendations, since they present different aspects of \cite{Zhao2013}, i.e., the used method or resource.
When considering documents connected to the seed (i.e., one common context) over two edges recommendations focusing on specific aspects are more feasible (either all green or all orange documents in Figure~\ref{fig:paper-example}). 
Diverse and focused recommendations could be especially suitable for scenarios in which the user's information need demands different perspectives for the same seed. 
In contrast to user-based recommender systems, content-based approaches usually struggle to account for specific preferences from their users. 
One way to respect different information requirements would be to suggest alternative recommendation sets that are focused on specific aspects. 
In the example of~\cite{Zhao2013}, focused recommendation sets could include paper using the same resource or applying the same method.
The intersection of contexts would even allow finding paper with the same resource but different methods.

Figure~\ref{fig:recsys-example} presents how contextual information could be integrated into a recommender system interface.
We use the research paper search engine Semantic Scholar\footnote{\url{https://www.semanticscholar.org/}} as example.
The three presented recommendations are diverse since they relate to different contexts (2) or a combination of contexts (3).
If users are interested in a particular aspect, they can click on the context icon to browse more focused recommendations.
The contextual document similarity is the foundation for such recommender  system.

\subsection{Preliminary Work} \label{ssec:preliminary-work}

We initiated our research on recommender systems with a large-scale offline evaluation of text- and link-based methods for the task of finding related Wikipedia articles~\cite{Schwarzer2016}\footnote{The papers \cite{Schwarzer2016,Schwarzer2017} are published under my birth name Malte Schwarzer.}. 
In~\cite{Schwarzer2016}, we demonstrate that the text-based TF-IDF~\cite{Jones1972} generates different recommendations compared to Co-citation~\cite{Small1973} and CPA~\cite{Gipp2009}.
Depending on individual information needs, one or the other recommender system is preferred (T1).
With Citolytics~\cite{Schwarzer2017}, we extend the concept of CPA and, moreover, show how a citation-based recommender system can be deployed into a production environment (T3).

A key contribution of this research is the hybrid use of content-based features (T2).
In~\cite{Ostendorff2019}, we incorporate text- and link-features for the document classification task.
We use contextual text representations from BERT~\cite{Devlin2019} and combine them with knowledge graph embeddings, i.e., vector representations derived from links between Wikidata items.
On a technical level, both vector representations, text and graph, are simply concatenated. 
Whether such a concatenation is the suitable method, is a question that we aim to answer with this research.

Recommender systems are only one example of applications that would benefit from contextual document similarity.
Another application is semantic storytelling, i.e., the semi-automatic arrangement of individual content pieces into a coherent story.
In \cite{Rehm2020c}, we took the first steps towards defining the similarity context as discourse relations between different text segments.
Our learnings from semantic storytelling facilitate the development of the prototypical recommender system (T3).

\subsection{Expected Contributions}
\label{ssec:objectives}

This doctoral thesis will make the following four key contributions.

\paragraph{Hybrid Text- and Link-based Document Representations} 

Text and links provide crucial semantic information about documents. 
While the document text reflects the content as expressed by its author, links reveal how a single document relates to the entire document collection.
We expect especially the textual context of a link as valuable information for determining the contextual similarity of linked documents.
Therefore, we will combine the two information sources for hybrid text- and link-based document representations (T2).
As Kanakia et al.~\cite{Kanakia2019} have shown, hybrid methods are beneficial for recommender systems.
Similarly, in \cite{Ostendorff2019}, we already combine text-based features with knowledge embeddings for document classification.
Our document representation technique will additionally incorporate the proximity of the link markers based on the idea of CPA~\cite{Gipp2009}.
Building upon existing graph embedding methods~\cite{Perozzi2014,Grover2016}, the link proximity can be integrated by formulating the citation network as a weighted graph.
To proof the validity of our hybrid technique, we will empirically evaluate it in comparison to well-established baselines, e.g., \cite{Jones1972,Le2014,Arora2017,Kanakia2019} and others.

\paragraph{Segment-level Document Similarity}

A simple approach for contextual document similarity is to split documents into semantic segments and to compute the similarity on the segment-level.
For specific document types, which have a standardized format, such an approach is applicable. 
Research papers can be segmented, for example, along their sections into background, methodology, and conclusion.
Chan et al.~\cite{Chan2018} present a proof-of-concept for this approach.
For other domains, the segmentation is less trivial, e.g., for non-standardized literature or when the segmentation cannot be not predefined.
In such settings, the challenge is to split a coherent document into segments automatically.
The document segmentation is, in its core, a sequence labeling task.
In~\cite{Schulz2020}, we already investigate methods, like BERT~\cite{Devlin2019}, for the labeling of named entities in medical documents.
As part of the doctoral thesis, we will continue this work for document segmentation and test the method against existing benchmarks~\cite{Chan2018}. 

\paragraph{Pairwise Document Classification}

The segmentation of documents is a practical and straightforward approach. 
However, the segmentation simultaneously eliminates the coherence of the document content.
One segment may lose its semantic meaning without the surrounding segments.  
We address this issue already in~\cite{Rehm2020c} in regards to coherent discourse relations. 
Consequently, the segment-level document similarity will only be a suboptimal solution.
A way to achieve context document similarity while keeping the document coherence intact is to treat the problem as pairwise document classification.
The document pair of $d_s$ and $d_t$ is subject to a classifier that predicts the class label $c_i$, whereby the prediction probability corresponds to the similarity score.
We demonstrate the validity of this approach in~\cite{Ostendorff2020}. 
Namely, we evaluate different classifiers based on GloVe~\cite{Pennington2014}, Paragraph Vectors~\cite{Le2014}, BERT~\cite{Devlin2019}, and XLNet~\cite{Yang2019} with a dataset constructed from Wikidata and Wikipedia. 
In the dataset, Wikipedia articles are the documents, and Wikidata statements and properties define how the documents relate to each other.
The work presented in~\cite{Ostendorff2020} can be considered as the first step in this research directory based on a tailored dataset.
Next, we will extend the pairwise document classification to other corpora. 
The primary goal is its application on academic literature, whereby we will utilize data from existing academic knowledge graphs~\cite{Jaradeh2019,Farber2019} but also manually annotate documents for this purpose.

\paragraph{User-centric Evaluation}

In addition to the offline and data-centric evaluation, a method that underpins recommender systems ultimately needs to be evaluated in terms of user satisfaction.
The contradictory results of past offline and online evaluations highlight this need~\cite{Beel2016b}.
As a result, the user-centric evaluation will be the fourth contribution of this thesis (T4).
On the one hand, we are currently exploring the perceived differences between text- and link-based similarity measures with a qualitative user study. 
The ongoing study will provide a better understanding of changing information needs.
We investigate what user properties affect the information need and what method addresses these needs the best.
Besides studying the information need, we also need to collect additional expert feedback. 
This feedback is a prerequisite before specifying the set of similar contexts for a literature domain.

On the other hand, we will conduct an online evaluation to counter the influence of a laboratory set-up in user studies.
Online evaluations are inherently challenging to conduct since they require access to an application with a sufficient number of users.
But when successfully conducted, online evaluations produce findings under realistic settings.
Additionally, online evaluations lead more quickly to statistically significant results compared to small-scale user studies.
Due to these reasons, we have invested special effort into establishing research partnerships with non-profit digital libraries for the purpose of online evaluations.
For research papers, we will collaborate with Mr. DLib~\cite{Beel2011}, while CourtListener~\cite{Lissner2010} and Open Legal Data~\cite{Ostendorff2020b} will provide us the opportunity to investigate legal documents.
Moreover, we will evaluate recommendations of mathematical literature with the support of zbMATH\footnote{\url{https://zbmath.org/}}.
Collaborating with non-profit partners will also contribute to the openness our research. 
For instance, we expect that experimental data can be published more easily without conflicting commercial interests.
Publishing experimental data is critical when working against the reproducibility issues in recommender system research~\cite{Beel2016b,Konstan2013}.
With this focus on transparent and reproducible evaluations, we emphasize these issues.

\section{Conclusion} \label{sec:conclusion}

In summary, literature recommender systems could benefit significantly from the development of contextual document similarity measures.
With the conceived system, users will be able to explore document collections by formulating queries in terms of documents and how they relate to each other.
To achieve for this goal, we will develop a prototype and evaluate it for different literature domains.

In contrast to existing document similarity measures that distinguish between similar or dissimilar, and fail to express what makes two documents alike,
the investigated measure will provide a context to the similarity.
The contextual document similarity is defined as a triple of two documents and the context that specifies to what the similarity of the two documents relates.
On a technical level, we will incorporate semantic features from text and links in a hybrid manner to represent documents.

\bibliography{paper}

\begin{thebibliography}{62}
\expandafter\ifx\csname natexlab\endcsname\relax\def\natexlab#1{#1}\fi

\bibitem[{Arora et~al.(2017)Arora, Liang, and Ma}]{Arora2017}
Sanjeev Arora, Yingyu Liang, and Tengyu Ma. 2017.
\newblock {A simple but though Baseline for Sentence Embeddings}.
\newblock In \emph{5th International Conference on Learning Representations
  (ICLR 2017)}, volume~15, pages 416--424.

\bibitem[{Athar and Teufel(2012)}]{Athar2012}
Awais Athar and Simone Teufel. 2012.
\newblock \href {https://doi.org/10.1002/asi} {{Detection of Implicit Citations
  for Sentiment Detection}}.
\newblock \emph{Proceedings of the Workshop on Detecting Structure in Scholarly
  Discourse}, (July):18--26.

\bibitem[{Bai et~al.(2019)Bai, Wang, Lee, Yang, Kong, and Xia}]{Bai2019}
Xiaomei Bai, Mengyang Wang, Ivan Lee, Zhuo Yang, Xiangjie Kong, and Feng Xia.
  2019.
\newblock \href {https://doi.org/10.1109/ACCESS.2018.2890388} {{Scientific
  paper recommendation: A survey}}.
\newblock \emph{IEEE Access}, 7:9324--9339.

\bibitem[{B{\"{a}}r et~al.(2011)B{\"{a}}r, Zesch, and Gurevych}]{Bar2011}
Daniel B{\"{a}}r, Torsten Zesch, and Iryna Gurevych. 2011.
\newblock {A reflective view on text similarity}.
\newblock \emph{International Conference Recent Advances in Natural Language
  Processing, RANLP}, (September):515--520.

\bibitem[{Beel et~al.(2016{\natexlab{a}})Beel, Breitinger, Langer, Lommatzsch,
  and Gipp}]{Beel2016b}
Joeran Beel, Corinna Breitinger, Stefan Langer, Andreas Lommatzsch, and Bela
  Gipp. 2016{\natexlab{a}}.
\newblock \href {https://doi.org/10.1007/s11257-016-9174-x} {{Towards
  reproducibility in recommender-systems research}}.
\newblock \emph{User Modeling and User-Adapted Interaction (UMAI)}, 26.

\bibitem[{Beel et~al.(2016{\natexlab{b}})Beel, Gipp, Langer, and
  Breitinger}]{Beel2016}
Joeran Beel, Bela Gipp, Stefan Langer, and Corinna Breitinger.
  2016{\natexlab{b}}.
\newblock \href {https://doi.org/10.1007/s00799-015-0156-0} {{Research-paper
  recommender systems: a literature survey}}.
\newblock \emph{International Journal on Digital Libraries}, 17(4):305--338.

\bibitem[{Beel et~al.(2011)Beel, Gipp, Langer, Genzmehr, Wilde,
  N{\"{u}}rnberger, and Pitman}]{Beel2011}
Joeran Beel, Bela Gipp, Stefan Langer, Marcel Genzmehr, Erik Wilde, Andreas
  N{\"{u}}rnberger, and Jim Pitman. 2011.
\newblock \href {https://doi.org/10.1145/1998076.1998187} {{Introducing Mr.
  DLib, a Machine-readable Digital Library}}.
\newblock In \emph{Proceeding of the 11th annual international ACM/IEEE joint
  conference on Digital libraries - JCDL '11}, page 463, New York, New York,
  USA. ACM Press.

\bibitem[{Berger et~al.(2017)Berger, McDonough, and Seversky}]{Berger2017}
Matthew Berger, Katherine McDonough, and Lee~M. Seversky. 2017.
\newblock \href {https://doi.org/10.1109/TVCG.2016.2598667} {{cite2vec:
  Citation-Driven Document Exploration via Word Embeddings}}.
\newblock \emph{IEEE Transactions on Visualization and Computer Graphics},
  23(1):691--700.

\bibitem[{Betts et~al.(2019)Betts, Power, and Ammar}]{Betts2019}
Christine Betts, Joanna Power, and Waleed Ammar. 2019.
\newblock \href {https://doi.org/10.18653/v1/P19-3025} {{GrapAL: Connecting the
  Dots in Scientific Literature}}.
\newblock In \emph{Proceedings of the 57th Annual Meeting of the Association
  for Computational Linguistics: System Demonstrations}, volume~13, pages
  147--152, Stroudsburg, PA, USA. Association for Computational Linguistics.

\bibitem[{Bojanowski et~al.(2017)Bojanowski, Grave, Joulin, and
  Mikolov}]{Bojanowski2017}
Piotr Bojanowski, Edouard Grave, Armand Joulin, and Tomas Mikolov. 2017.
\newblock \href {http://arxiv.org/abs/1607.04606} {{Enriching Word Vectors with
  Subword Information}}.
\newblock \emph{Transactions of the Association for Computational Linguistics},
  5:135--146.

\bibitem[{Chan et~al.(2018)Chan, Chang, Hope, Shahaf, and Kittur}]{Chan2018}
Joel Chan, Joseph~Chee Chang, Tom Hope, Dafna Shahaf, and Aniket Kittur. 2018.
\newblock \href {https://doi.org/10.1145/3274300} {{SOLVENT: A Mixed Initiative
  System for Finding Analogies between Research Papers}}.
\newblock \emph{Proceedings of the ACM on Human-Computer Interaction},
  2(CSCW):1--21.

\bibitem[{Cohan et~al.(2019)Cohan, Ammar, van Zuylen, and Cady}]{Cohan2019}
Arman Cohan, Waleed Ammar, Madeleine van Zuylen, and Field Cady. 2019.
\newblock \href {https://doi.org/10.18653/v1/N19-1361} {{Structural Scaffolds
  for Citation Intent Classification in Scientific Publications}}.
\newblock In \emph{Proceedings of the 2019 Conference of the North}, volume~1,
  pages 3586--3596, Stroudsburg, PA, USA. Association for Computational
  Linguistics.

\bibitem[{Collins and Beel(2019)}]{Collins2019}
Andrew Collins and Joeran Beel. 2019.
\newblock \href {http://arxiv.org/abs/1905.11244} {{Document Embeddings vs.
  Keyphrases vs. Terms: An Online Evaluation in Digital Library Recommender
  Systems}}.
\newblock In \emph{ACM/IEEE Joint Conference on Digital Libraries (JCDL)},
  pages 130--133.

\bibitem[{Cortes and Vapnik(1995)}]{Cortes1995}
Corinna Cortes and Vladimir Vapnik. 1995.
\newblock \href {https://doi.org/10.1007/BF00994018} {{Support-vector
  networks}}.
\newblock \emph{Machine Learning}, 20(3):273--297.

\bibitem[{Dai et~al.(2019)Dai, Yang, Yang, Carbonell, Le, and
  Salakhutdinov}]{Dai2019}
Zihang Dai, Zhilin Yang, Yiming Yang, Jaime Carbonell, Quoc Le, and Ruslan
  Salakhutdinov. 2019.
\newblock \href {https://doi.org/10.18653/v1/P19-1285} {{Transformer-XL:
  Attentive Language Models beyond a Fixed-Length Context}}.
\newblock In \emph{Proceedings of the 57th Annual Meeting of the Association
  for Computational Linguistics}, pages 2978--2988, Stroudsburg, PA, USA.
  Association for Computational Linguistics.

\bibitem[{Devlin et~al.(2019)Devlin, Chang, Lee, and Toutanova}]{Devlin2019}
Jacob Devlin, Ming-Wei Chang, Kenton Lee, and Kristina Toutanova. 2019.
\newblock \href {https://doi.org/10.18653/v1/N19-1423} {{BERT: Pre-training of
  Deep Bidirectional Transformers for Language Understanding}}.
\newblock In \emph{Proceedings of the 2019 Conference of the North American
  Chapter of the Association for Computational Linguistics}, pages 4171--4186,
  Minneapolis, Minnesota. Association for Computational Linguistics.

\bibitem[{Elkiss et~al.(2008)Elkiss, Shen, Fader, Erkan, States, and
  Radev}]{Elkiss2008}
Aaron Elkiss, Siwei Shen, Anthony Fader, G{\"{u}}neş Erkan, David States, and
  Dragomir Radev. 2008.
\newblock \href {https://doi.org/10.1002/asi.20707} {{Blind men and elephants:
  What do citation summaries tell us about a research article?}}
\newblock \emph{Journal of the American Society for Information Science and
  Technology}, 59(1):51--62.

\bibitem[{Eto(2013)}]{Eto2013}
Masaki Eto. 2013.
\newblock \href {https://doi.org/10.1007/s11192-012-0756-z} {{Evaluations of
  context-based co-citation searching}}.
\newblock \emph{Scientometrics}, 94(2):651--673.

\bibitem[{F{\"{a}}rber(2019)}]{Farber2019}
Michael F{\"{a}}rber. 2019.
\newblock \href {https://doi.org/10.1007/978-3-030-30796-7_8} {{The Microsoft
  Academic Knowledge Graph: A Linked Data Source with 8 Billion Triples of
  Scholarly Data}}.
\newblock In \emph{Proceedings of the 18th International Semantic Web
  Conference}, pages 113--129.

\bibitem[{Ferrara et~al.(2011)Ferrara, Pudota, and Tasso}]{Ferrara2011}
Felice Ferrara, Nirmala Pudota, and Carlo Tasso. 2011.
\newblock \href {https://doi.org/10.1007/978-3-642-27302-5_2} {{A
  keyphrase-based paper recommender system}}.
\newblock \emph{Communications in Computer and Information Science}, 249
  CCIS:14--25.

\bibitem[{Ganguly and Pudi(2017)}]{Ganguly2017}
Soumyajit Ganguly and Vikram Pudi. 2017.
\newblock \href {https://doi.org/10.1007/978-3-319-56608-5_30} {{Paper2vec:
  Combining graph and text information for scientific paper representation}}.
\newblock \emph{Lecture Notes in Computer Science (including subseries Lecture
  Notes in Artificial Intelligence and Lecture Notes in Bioinformatics)}, 10193
  LNCS:383--395.

\bibitem[{Garfield(2001)}]{Garfield2001}
Eugene Garfield. 2001.
\newblock {From Bibliographic Coupling to Co-Citation Analysis Via Algorithmic
  Historio-Bibliography}.

\bibitem[{Gick and Holyoak(1983)}]{Gick1983}
Mary~L. Gick and Keith~J. Holyoak. 1983.
\newblock \href {https://doi.org/10.1016/0010-0285(83)90002-6} {{Schema
  induction and analogical transfer}}.
\newblock \emph{Cognitive Psychology}, 15(1):1--38.

\bibitem[{Gipp and Beel(2009)}]{Gipp2009}
Bela Gipp and J{\"{o}}ran Beel. 2009.
\newblock \href {https://doi.org/10.1045/november2009-inbrief.URL} {{Citation
  Proximity Analysis ( CPA ) – A new approach for identifying related work
  based on Co-Citation Analysis}}.
\newblock \emph{ISSI '09: Proceedings of the 12th International Conference on
  Scientometrics and Informetrics}, 2(July):571--575.

\bibitem[{Goodman(1972)}]{Goodman1972}
N~Goodman. 1972.
\newblock {Seven strictures on similarity}.
\newblock \emph{Problems and projects}.

\bibitem[{Grover and Leskovec(2016)}]{Grover2016}
Aditya Grover and Jure Leskovec. 2016.
\newblock \href {https://doi.org/10.1145/2939672.2939754} {{node2vec: Scalable
  Feature Learning for Networks}}.
\newblock In \emph{Proceedings of the 22nd ACM SIGKDD International Conference
  on Knowledge Discovery and Data Mining - KDD '16}, pages 855--864, New York,
  New York, USA. ACM Press.

\bibitem[{Han et~al.(2018)Han, Song, Zhao, Shi, and Zhang}]{Han2018}
Jialong Han, Yan Song, Wayne~Xin Zhao, Shuming Shi, and Haisong Zhang. 2018.
\newblock \href {https://doi.org/10.18653/v1/P18-1222} {{hyperdoc2vec:
  Distributed Representations of Hypertext Documents}}.
\newblock In \emph{Proceedings of the 56th Annual Meeting of the Association
  for Computational Linguistics (Volume 1: Long Papers)}, volume~1, pages
  2384--2394, Stroudsburg, PA, USA. Association for Computational Linguistics.

\bibitem[{Harris(1954)}]{Harris1954}
Zellig~S. Harris. 1954.
\newblock \href {https://doi.org/10.1080/00437956.1954.11659520}
  {{Distributional Structure}}.
\newblock \emph{WORD}, 10(2-3):146--162.

\bibitem[{Iacobacci et~al.(2016)Iacobacci, Pilehvar, and
  Navigli}]{Iacobacci2016}
Ignacio Iacobacci, Mohammad~Taher Pilehvar, and Roberto Navigli. 2016.
\newblock \href {https://doi.org/10.18653/v1/p16-1085} {{Embeddings for word
  sense disambiguation: An evaluation study}}.
\newblock \emph{54th Annual Meeting of the Association for Computational
  Linguistics, ACL 2016 - Long Papers}, 2(2003):897--907.

\bibitem[{Jannach et~al.(2010)Jannach, Zanker, Felfernig, and
  Friedrich}]{Jannach2010}
Dietmar Jannach, Markus Zanker, Alexander Felfernig, and Gerhard Friedrich.
  2010.
\newblock \emph{{Recommender Systems - An Introduction}}, 1 edition.
\newblock Cambridge University Press, United Kingdom.

\bibitem[{Jaradeh et~al.(2019)Jaradeh, Oelen, Farfar, Prinz, D'Souza,
  Kismih{\'{o}}k, Stocker, and Auer}]{Jaradeh2019}
Mohamad~Yaser Jaradeh, Allard Oelen, Kheir~Eddine Farfar, Manuel Prinz,
  Jennifer D'Souza, G{\'{a}}bor Kismih{\'{o}}k, Markus Stocker, and S{\"{o}}ren
  Auer. 2019.
\newblock \href {https://doi.org/10.1145/3360901.3364435} {{Open Research
  Knowledge Graph: Next Generation Infrastructure for Semantic Scholarly
  Knowledge}}.
\newblock pages 243--246.

\bibitem[{Jones(1972)}]{Jones1972}
Karen~Sparck Jones. 1972.
\newblock \href {https://doi.org/10.1108/eb026526} {{A statistical
  interpretation of term specificity and its application in retrieval}}.
\newblock \emph{Journal of Documentation}, 28(1):11--21.

\bibitem[{Jurgens et~al.(2018)Jurgens, Kumar, Hoover, McFarland, and
  Jurafsky}]{Jurgens2018}
David Jurgens, Srijan Kumar, Raine Hoover, Dan McFarland, and Dan Jurafsky.
  2018.
\newblock \href {https://doi.org/10.1162/tacl_a_00028} {{Measuring the
  Evolution of a Scientific Field through Citation Frames}}.
\newblock \emph{Transactions of the Association for Computational Linguistics},
  6:391--406.

\bibitem[{Kanakia et~al.(2019)Kanakia, Shen, Eide, and Wang}]{Kanakia2019}
Anshul Kanakia, Zhihong Shen, Darrin Eide, and Kuansan Wang. 2019.
\newblock \href {https://doi.org/10.1145/3308558.3313700} {{A Scalable Hybrid
  Research Paper Recommender System for Microsoft Academic}}.
\newblock In \emph{The World Wide Web Conference on - WWW '19}, pages
  2893--2899, New York, New York, USA. ACM Press.

\bibitem[{Kessler(1963)}]{Kessler1963}
M.~M. Kessler. 1963.
\newblock \href {https://doi.org/10.1002/asi.5090140103} {{Bibliographic
  coupling between scientific papers}}.
\newblock \emph{American Documentation}, 14(1):10--25.

\bibitem[{Knoth and Khadka(2017)}]{Knoth2017}
Petr Knoth and Anita Khadka. 2017.
\newblock {Can we do better than Co-Citations? - Bringing Citation Proximity
  Analysis from idea to practice in research article recommendation}.
\newblock In \emph{2nd Joint Workshop on Bibliometric-enhanced Information
  Retrieval and Natural Language Processing for Digital Libraries}, Tokyo,
  Japan.

\bibitem[{Konstan and Adomavicius(2013)}]{Konstan2013}
Joseph~A. Konstan and Gediminas Adomavicius. 2013.
\newblock \href {https://doi.org/10.1145/2532508.2532513} {{Toward
  identification and adoption of best practices in algorithmic recommender
  systems research}}.
\newblock In \emph{Proceedings of the International Workshop on Reproducibility
  and Replication in Recommender Systems Evaluation - RepSys '13}, volume
  2532513, pages 23--28, New York, New York, USA. ACM Press.

\bibitem[{Lau and Baldwin(2016)}]{Lau2016}
Jey~Han Lau and Timothy Baldwin. 2016.
\newblock \href {https://doi.org/10.18653/v1/W16-1609} {{An Empirical
  Evaluation of doc2vec with Practical Insights into Document Embedding
  Generation}}.
\newblock In \emph{Proceedings of the 1st Workshop on Representation Learning
  for NLP}, pages 78--86, Stroudsburg, PA, USA. Association for Computational
  Linguistics.

\bibitem[{Le and Mikolov(2014)}]{Le2014}
Quoc~V. Le and Tomas Mikolov. 2014.
\newblock \href {http://arxiv.org/abs/1405.4053} {{Distributed Representations
  of Sentences and Documents}}.
\newblock \emph{International conference on machine learning}, 32:1188--1196.

\bibitem[{Lissner(2010)}]{Lissner2010}
Michael Lissner. 2010.
\newblock \href
  {https://www.ischool.berkeley.edu/files/student{\_}projects/Final{\_}Report{\_}Michael{\_}Lissner{\_}2010-05-07{\_}2.pdf}
  {\emph{{CourtListener.com: A platform for researching and staying abreast of
  the latest in the law}}}.
\newblock Master thesis, University of California, Berkeley.

\bibitem[{Liu et~al.(2019)Liu, Ott, Goyal, Du, Joshi, Chen, Levy, Lewis,
  Zettlemoyer, and Stoyanov}]{Liu2019}
Yinhan Liu, Myle Ott, Naman Goyal, Jingfei Du, Mandar Joshi, Danqi Chen, Omer
  Levy, Mike Lewis, Luke Zettlemoyer, and Veselin Stoyanov. 2019.
\newblock \href {http://arxiv.org/abs/1907.11692} {{RoBERTa: A Robustly
  Optimized BERT Pretraining Approach}}.
\newblock (1).

\bibitem[{Mikolov et~al.(2013)Mikolov, Chen, Corrado, and Dean}]{Mikolov2013}
Tomas Mikolov, Kai Chen, Greg Corrado, and Jeffrey Dean. 2013.
\newblock \href {http://arxiv.org/abs/1301.3781} {{Efficient Estimation of Word
  Representations in Vector Space}}.
\newblock pages 1--12.

\bibitem[{{Mohamed Hassan} et~al.(2019){Mohamed Hassan}, Sansonetti,
  Gasparetti, Micarelli, and Beel}]{Hassan2019}
Hebatallah~A. {Mohamed Hassan}, Giuseppe Sansonetti, Fabio Gasparetti,
  Alessandro Micarelli, and Joeran Beel. 2019.
\newblock {BERT, ELMo, use and infersent sentence encoders: The Panacea for
  research-paper recommendation?}
\newblock In \emph{CEUR Workshop Proceedings}, volume 2431, pages 6--10.

\bibitem[{Nakov et~al.(2004)Nakov, Schwartz, and Hearst}]{Nakov2004}
Preslav~I Nakov, Ariel~S Schwartz, and Marti Hearst. 2004.
\newblock {Citances: Citation sentences for semantic analysis of bioscience
  text}.
\newblock \emph{Proceedings of the SIGIR'04 workshop on Search and Discovery in
  Bioinformatics}.

\bibitem[{Ostendorff et~al.(2020{\natexlab{a}})Ostendorff, Blume, and
  Ostendorff}]{Ostendorff2020b}
Malte Ostendorff, Till Blume, and Saskia Ostendorff. 2020{\natexlab{a}}.
\newblock \href {https://doi.org/10.1145/3383583.3398616} {{Towards an Open
  Platform for Legal Information}}.
\newblock In \emph{Proceedings of the 20th ACM/IEEE Joint Conference on Digital
  Libraries (JCDL`20)}.

\bibitem[{Ostendorff et~al.(2019)Ostendorff, Bourgonje, Berger,
  Moreno-Schneider, and Rehm}]{Ostendorff2019}
Malte Ostendorff, Peter Bourgonje, Maria Berger, Julian Moreno-Schneider, and
  Georg Rehm. 2019.
\newblock {Enriching BERT with Knowledge Graph Embedding for Document
  Classification}.
\newblock In \emph{Proceedings of the GermEval 2019 Workshop}, Erlangen,
  Germany.

\bibitem[{Ostendorff et~al.(2020{\natexlab{b}})Ostendorff, Ruas, Schubotz,
  Rehm, and Gipp}]{Ostendorff2020}
Malte Ostendorff, Terry Ruas, Moritz Schubotz, Georg Rehm, and Bela Gipp.
  2020{\natexlab{b}}.
\newblock \href {https://doi.org/10.1145/3383583.3398525} {{Pairwise
  Multi-Class Document Classification for Semantic Relations between Wikipedia
  Articles}}.
\newblock In \emph{Proceedings of the 20th ACM/IEEE Joint Conference on Digital
  Libraries (JCDL`20)}.

\bibitem[{Pennington et~al.(2014)Pennington, Socher, and
  Manning}]{Pennington2014}
Jeffrey Pennington, Richard Socher, and Christopher Manning. 2014.
\newblock \href {https://doi.org/10.3115/v1/D14-1162} {{Glove: Global Vectors
  for Word Representation}}.
\newblock In \emph{Proceedings of the 2014 Conference on Empirical Methods in
  Natural Language Processing (EMNLP)}, pages 1532--1543, Stroudsburg, PA, USA.
  Association for Computational Linguistics.

\bibitem[{Perozzi et~al.(2014)Perozzi, Al-Rfou, and Skiena}]{Perozzi2014}
Bryan Perozzi, Rami Al-Rfou, and Steven Skiena. 2014.
\newblock \href {https://doi.org/10.1145/2623330.2623732} {{DeepWalk: online
  learning of social representations}}.
\newblock In \emph{Proceedings of the 20th ACM SIGKDD international conference
  on Knowledge discovery and data mining - KDD '14}, pages 701--710, New York,
  New York, USA. ACM Press.

\bibitem[{Rehm et~al.(2020)Rehm, Zaczynska, Schneider, Ostendorff, Bourgonje,
  Berger, Rauenbusch, Schmidt, and Wild}]{Rehm2020c}
Georg Rehm, Karolina Zaczynska, Julian~Moreno Schneider, Malte Ostendorff,
  Peter Bourgonje, Maria Berger, Jens Rauenbusch, Andre Schmidt, and Mikka
  Wild. 2020.
\newblock {Towards Discourse Parsing-inspired Semantic Storytelling}.
\newblock In \emph{Proceedings of QURATOR 2020 -- The conference for
  intelligent content solutions}, Berin, Germany.

\bibitem[{Reimers and Gurevych(2019)}]{Reimers2019}
Nils Reimers and Iryna Gurevych. 2019.
\newblock \href {http://arxiv.org/abs/1908.10084} {{Sentence-BERT: Sentence
  Embeddings using Siamese BERT-Networks}}.
\newblock In \emph{The 2019 Conference on Empirical Methods in Natural Language
  Processing (EMNLP 2019)}.

\bibitem[{Ruas et~al.(2019)Ruas, Grosky, and Aizawa}]{Ruas2019}
Terry Ruas, William Grosky, and Akiko Aizawa. 2019.
\newblock \href {https://doi.org/10.1016/j.eswa.2019.06.026} {{Multi-sense
  embeddings through a word sense disambiguation process}}.
\newblock \emph{Expert Systems with Applications}, 136:288--303.

\bibitem[{Salton et~al.(1975)Salton, Wong, and Yang}]{Salton1975}
G.~Salton, A.~Wong, and C.~S. Yang. 1975.
\newblock {Vector Space Model for Automatic Indexing. Information Retrieval and
  Language Processing}.
\newblock \emph{Communications of the ACM}, 18(11):613--620.

\bibitem[{Schafer et~al.(2007)Schafer, Frankowski, Herlocker, and
  Sen}]{Schafer2007}
J~Ben Schafer, Dan Frankowski, Jon Herlocker, and Shilad Sen. 2007.
\newblock \href {https://doi.org/10.1007/978-3-540-72079-9_9} {{Collaborative
  Filtering Recommender Systems}}.
\newblock In \emph{The Adaptive Web}, volume 4321, pages 291--324. Springer
  Berlin Heidelberg, Berlin, Heidelberg.

\bibitem[{Schulz et~al.(2020)Schulz, Seva, Rodriguez, Ostendorff, and
  Rehm}]{Schulz2020}
Sarah Schulz, Jurica Seva, Samuel Rodriguez, Malte Ostendorff, and Georg Rehm.
  2020.
\newblock {Named Entities in Medical Case Reports: Corpus and Experiments}.
\newblock In \emph{Proceedings of the 12th International Conference on Language
  Resources and Evaluation}.

\bibitem[{Schwarzer et~al.(2017)Schwarzer, Breitinger, Schubotz, Meuschke, and
  Gipp}]{Schwarzer2017}
Malte Schwarzer, Corinna Breitinger, Moritz Schubotz, Norman Meuschke, and Bela
  Gipp. 2017.
\newblock \href {https://doi.org/10.1145/3109859.3109981} {{Citolytics: A
  Link-based Recommender System for Wikipedia}}.
\newblock In \emph{Proceedings of the Eleventh ACM Conference on Recommender
  Systems - RecSys '17}, pages 360--361, New York, New York, USA. ACM, ACM
  Press.

\bibitem[{Schwarzer et~al.(2016)Schwarzer, Schubotz, Meuschke, and
  Breitinger}]{Schwarzer2016}
Malte Schwarzer, Moritz Schubotz, Norman Meuschke, and Corinna Breitinger.
  2016.
\newblock \href {https://doi.org/10.1145/2910896.2910908} {{Evaluating
  Link-based Recommendations for Wikipedia}}.
\newblock \emph{Proceedings of the 16th ACM/IEEE Joint Conference on Digital
  Libraries (JCDL`16)}, pages 191--200.

\bibitem[{Small(1973)}]{Small1973}
Henry Small. 1973.
\newblock {A New Measure of the Relationship Two Documents}.
\newblock \emph{Journal of the American Society for Information Science}, 24.

\bibitem[{Vaswani et~al.(2017)Vaswani, Shazeer, Parmar, Uszkoreit, Jones,
  Gomez, Kaiser, and Polosukhin}]{Vaswani2017}
Ashish Vaswani, Noam Shazeer, Niki Parmar, Jakob Uszkoreit, Llion Jones,
  Aidan~N. Gomez, Lukasz Kaiser, and Illia Polosukhin. 2017.
\newblock \href {http://arxiv.org/abs/1706.03762} {{Attention Is All You
  Need}}.
\newblock \emph{Proceedings of the 31st International Conference on Neural
  Information Processing Systems}, (Nips):6000--6010.

\bibitem[{Yang et~al.(2019)Yang, Dai, Yang, Carbonell, Salakhutdinov, and
  Le}]{Yang2019}
Zhilin Yang, Zihang Dai, Yiming Yang, Jaime Carbonell, Ruslan Salakhutdinov,
  and Quoc~V Le. 2019.
\newblock \href {http://arxiv.org/abs/1906.08237v1} {{XLNet: Generalized
  Autoregressive Pretraining for Language Understanding}}.
\newblock In \emph{Advances in Neural Information Processing Systems 32}, pages
  5754--5764.

\bibitem[{Zhao et~al.(2013)Zhao, Wang, and Huang}]{Zhao2013}
Jianyu Zhao, Peng Wang, and Kai Huang. 2013.
\newblock \href {https://doi.org/10.1145/2517288.2517298} {{A semi-supervised
  approach for author disambiguation in KDD CUP 2013}}.
\newblock In \emph{Proceedings of the 2013 KDD Cup 2013 Workshop on - KDD Cup
  '13}, pages 1--8, New York, New York, USA. ACM Press.

\bibitem[{Zheng et~al.(2010)Zheng, Wang, Zhang, Li, and Yang}]{Zheng2010}
Hua Zheng, Dong Wang, Qi~Zhang, Hang Li, and Tinghao Yang. 2010.
\newblock \href {https://doi.org/10.1145/1864708.1864759} {{Do clicks measure
  recommendation relevancy?}}
\newblock In \emph{Proceedings of the fourth ACM conference on Recommender
  systems - RecSys '10}, page 249, New York, New York, USA. ACM Press.

\end{thebibliography}
\bibliographystyle{acl_natbib}

\end{document}